# Surface and edge resonances of phonon-polaritons in scattering-type near-field optical microscopy

V.E. Babicheva

*Center for Nano-Optics, Georgia State University, Atlanta, GA, USA*
*email:* baviev@gmail.com

**Abstract**. We theoretically study resonance responses of flat surfaces and sharp edges of the nanostructures that support excitations of phonon-polaritons in mid-infrared range. We focus on two materials: silicon carbide that has a nearly isotropic permittivity and hexagonal boron nitride that has a strong anisotropy and spectral band with hyperbolic dispersion. We aim to predict scattering-type near-field optical microscope (s-SNOM) response and develop a modeling approach that adequately describes the resonant behavior of the nanostructure with phonon-polaritons. The previously employed technique assumes dipole scattering from the tip and allows calculating s-SNOM signal in different demodulation orders by modeling full structure, any tip positions, and vertical scans, which works well for the structures with only one hot spot, e.g. flat surfaces. In the structures of complex shapes, hot-spot places are unknown, and analysis of light absorption in the whole apex is the best way to account for all hot spots and field enhancement. We show that calculation of demodulation orders of light absorption in the tip is an alternative way to predict s-SNOM signal, and it is preferred for the structures of complex shapes with strong resonances, where dipole approximation of the tip is not valid.

**Keywords**: mid-infrared, silicon carbide, polar dielectric, hexagonal boron nitride, hyperbolic dispersion, phonon polaritons, near-field, optical resonances

**I. Introduction**

Recently emerged materials that support excitation of phonon-polaritons with low optical losses have attracted a lot of interest [1-11]. Non-radiative losses are relatively small for phonon-polariton excitations in silicon carbide (SiC) and hexagonal boron nitride (hBN) in comparison to plasmon resonances in metals. Silicon carbide is a polar dielectric and has negative permittivity around the wavelength 11 µm [6-8]. Furthermore, hBN is a van der Waals material and has a strong optical anisotropy [2,10,11]. In-plane or out-of-plane tensor components of permittivity can be negative at the particular spectral range, which is usually referred as two Reststrahlen bands: approximately around wavelength 7.3 µm (1370 cm$^{-1}$, the upper band) and 12.8 µm (780 cm$^{-1}$, the lower band). We are interested in the region of $\lambda \approx 6.2 - 7.3$ µm, where hexagonal boron nitride has type II hyperbolic dispersion [12], meaning in-plane components are negative and out-of-plane component is positive. hBN is a subject of active study because of its low losses and natural hyperbolicity, rarely available in nature [13]. Instead of designing a complex structure with alternating metal and dielectric layers of precise thickness [14-17] or growing nanowire arrays [12], one can use layers of hBN to confine the light and design optical elements such as nanoantennas [18,19], hyperlenses [20,21], waveguides [22-25], tapers [26], etc.

Scattering-type scanning near-field optical microscopy (s-SNOM) enables spatial resolution several orders of magnitude below the diffraction limit and has been shown to be effective for optical characterizations of different material surfaces. For example, s-SNOM can be applied to characterize material properties, analyze permittivity of the samples [27-30] and phase segregation [31,32], distinguish dielectric and metal nanoparticles [33], image surface waves [34-36], etc. One of the most interesting applications of s-SNOM is to study resonance excitations in the structures, such as plasmon- or phonon-polariton resonances. High-quality resonances in the structure can be accompanied by increase or decrease of the s-SNOM signal, and the image features are highly affected by the presence of the tip itself as it is the main scattering element in the system. The images successfully reproduce features of the structures with subwavelength resolution and field enhancement in their particular parts (e.g. hot spots or particle modes), but theoretical modeling of these processes is still limited by the dipole approximation of the tip scattering [37,38]. Quantitative description of the s-SNOM results and their correlation with the field enhancement in the structure are under active study [39-41], including recently reported technique allowing to reconstruct vertical interaction [41]. We have previously developed an approach that combines full-wave numerical simulations of tip-sample near-field interaction and signal demodulation at higher orders in accordance with typical s-SNOM experiments [32]. Unlike previous models, our technique allows modeling of the realistic tip and sample structure, but signal calculations are still limited by dipole approximation of the tip-scattered wave.

In the present work, we make the next step towards theoretical prediction of s-SNOM response. We analyze the link between near-fields enhancement in the tip-sample structure and s-SNOM measurements and directly relate them to changes in the effective polarizability of the tip. In particular, we study s-SNOM response for the structures that support phonon-polariton resonant excitations: first, structures where conditions of the resonance are well-known, such as SiC flat surface and nanospheres; second, an edge of SiC, which we consider an isotropic material with negative permittivity because of the phonon-polariton resonances; and finally, an edge of material with hyperbolic dispersion such as hBN with phonon-polariton rays and their multiple reflection from the material boundaries. The outside bright fringes correspond to the most efficient excitation of edge resonances, and we study surface and edge resonances as an example where s-SNOM signal show additional peaks (outside bright fringes and low-contrast lines appear in the measurements [29,30]). We find the connection between effective polarizability of the tip and enhancement of the field in different parts of the structure. Outside the resonance conditions, where only strong reflection from the structure occurs, the effective polarizability of the tip is increased and fields are enhanced. The behavior is the opposite in the resonant structure: the increase of fields is accompanied by the decrease of effective polarizability. Thus, we go beyond dipole model by calculating demodulation orders of absorption in the tip, which is the best way to account for all hot spots and changes of scattering from the tip.

**II. Field analysis**

We perform full-wave numerical modeling of the structures with realistic parameters: sizes, shapes, permittivities, and substrates (see Appendix for the details), and throughout the paper, we work in p-polarization. We compare the following quantities obtained from full-wave simulations and their dependencies from either vertical $z_{tip}$- or horizontal $x_{tip}$- position of the tip: (i) effective polarizability of the tip obtained from reflectance of the structure considering dipole approximation; (ii) s-SNOM signal calculated from effective polarizability of the tip; (iii) energy absorption (or non-radiative loss) in the apex of the tip (bottom hemisphere) and its demodulation orders ($2^{nd}$, $3^{rd}$, and $4^{th}$ harmonics); and (iv) electric field (or field enhancement) in different parts of the structure.

We extract effective polarizability of the tip through reflectance calculations of the domain and then calculate s-SNOM signal (see Appendix). From this modeling, we can also extract other parameters of interest, such as the electric field at any part of the tip or structure and, more generally, non-radiative energy loss in some parts of the structure. The latter one is useful because we collect a general information about the field in the structure if we do not know where the hot spots are excited and how many of them there are (edge characterization involves at least two hot spots). Absorption is ~$E^2$, and any changes in its total value (integral over tip apex) indicate the accounts for all hot spots in the structure. Similar to the calculations of the s-SNOM signal based on effective polarizability at different vertical positions $z_{tip}$, we find demodulation orders of the absorption based on absorption calculations at different $z_{tip}$. Thus, the main advantage of this approach is that we are not limited by the dipole approximation and its calculations based on reflectance, but rather account for all enhanced fields in the combined tip-sample structure. Below, we compare results obtained from both effective-polarizability and absorption calculations.

**III. Surface resonances: Effective polarizability of the tip and field enhancement**

We begin our study with the analysis of the simplest structures that either supporting or not supporting resonances (Fig. 1). The typical non-resonant structure is highly reflective surface e.g. gold in mid-infrared range (at λ = 10.55 µm its $\varepsilon_{Au}$ = -5100 + 970i, data from [42]), which in general resembles properties of perfect electric conductor possessing effectively infinite conductivity. Numerical modeling shows that effective polarizability of the tip on top of such non-resonant surface monotonically decreases with an increase of vertical position $z_{tip}$ (Fig. 1b). The same monotonic decrease is observed in other quantities, such as absorption in the tip and field enhancement in the tip (not shown here), which eventually results in the same behavior of s-SNOM signal in the experimental measurements (approach curve).

Next, we perform calculations for surfaces with resonances, and below we show that their characterization signatures are essentially different. Surface with negative permittivity $\varepsilon_m$ in the air supports propagating bound waves with propagation constant $k_{sphp} = (\omega/c)\sqrt{\varepsilon_m/(\varepsilon_m + 1)}$ (ω is frequency and *c* is the speed of light), and thus possesses resonant properties at Re($\varepsilon_m$) = -1. We study SiC surface at the wavelength 10.6 µm where its $\varepsilon_m$ = -1 + 0.08i (Fig. 1). We emphasize it is not simply a regime of surface phonon-polariton (SPhP) excitation and propagation, it is the resonance of SPhPs because of the close-to-zero denominator of propagation constant $k_{sphp}$. Note this SPhP resonance still has a finite field enhancement because of the damping provided by the non-zero imaginary part of permittivity. For $z_{tip}$ < 300 nm, the effective polarizability of the tip and absorption in the tip exhibit the changes opposite to non-resonant surface: both effective polarizability and absorption monotonically increase. For $z_{tip}$ > 300 nm, effective polarizability decreases similar to the non-resonant surface (outside the plot range). In contrast, absorption in the part of the resonant surface is maximal once the tip is close to the surface. These calculations confirm that upon excitation of resonance on the surface, scattering from the tip significantly decreases, which results in a decrease of effective polarizability.

This conclusion is also supported by the calculations of the resonant sphere (*R* = 30 nm) with localized surface phonon resonance (LSPhR). Field enhancement in the sphere in quasi-static approximation is defined as $\varepsilon_m/(\varepsilon_m + 2)$, and thus the resonance condition can be achieved at Re($\varepsilon_m$) = -2. The hypothetical spherical particle is chosen to apply the well-known analytical solution, and we model ideal sphere with $\varepsilon_m$ = -2 + 0.11i, which corresponds to SiC at wavelength 10.7 µm, located on the silicon substrate. Numerical simulations show non-monotonic changes in effective polarizability of the tip and absorptions, as well as the opposite tendencies in absorption in the tip and in the sphere. These changes happen within the smaller distance between tip and surface for the sphere than for the flat surface. It can be explained by the fact that the field enhancement is highly localized within the sphere while the flat surface is extended to the long distance, which becomes apparent comparing field distributions for sphere and flat surface (Fig. 1 c and f).

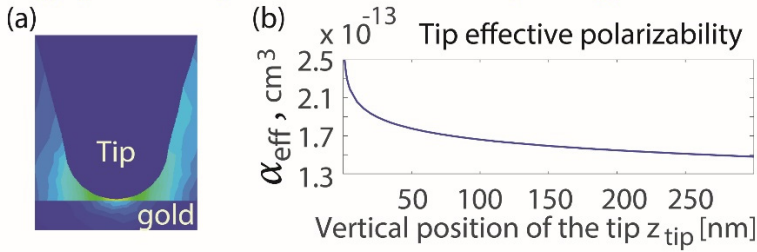

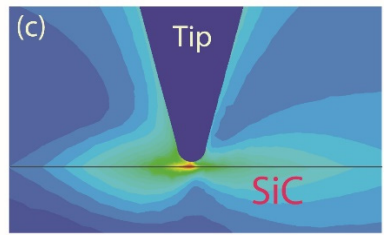
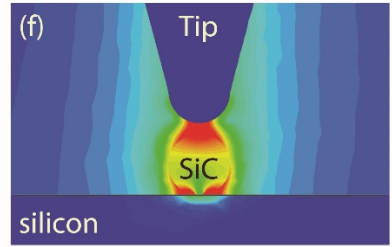

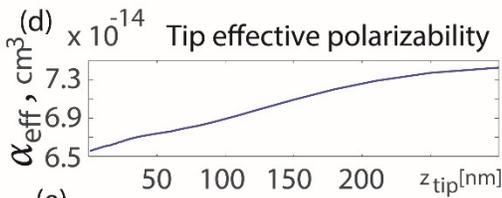
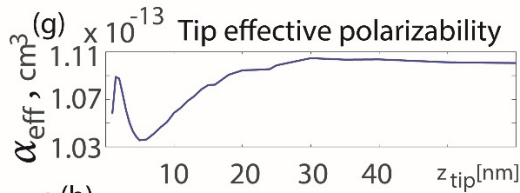

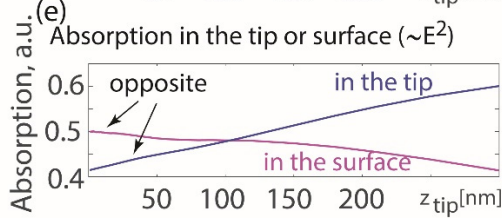
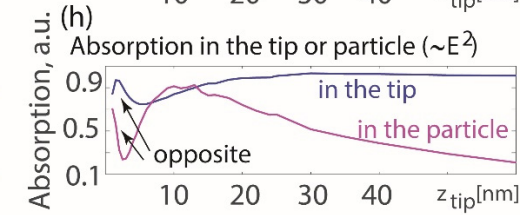

Fig. 1. Resonant and non-resonant surfaces. (a) s-SNOM tip above highly-reflective non-resonant surface (gold in mid-infrared range); (b) Change of effective polarizability of the tip moving away from gold surface: monotonic decrease is consistent with typical approach-curve measurements; (c) Tip and field distributions for the case of surface supporting SPhP resonance (single interface of SiC with $\varepsilon_m = -1 + 0.08i$ and air); (d) Effective polarizability of the tip moving away from the resonant surface: monotonic increase for $z_{tip}$ < 300 nm followed by the decrease (not shown in the plot); (e) Change of absorption in the apex of the tip (bottom hemisphere) and resonant surface (0.2x0.2x0.2 μm³ part): Absorption in the surface shows the *opposite* tendency than absorption in the tip and its effective polarizability. (f) Tip and field distributions for the case of sphere with localized surface phonon resonance ($\varepsilon_m = -2 + 0.11i$, corresponds to resonance in sphere, see equation in the text); (g) Effective polarizability of the tip moving away from resonant sphere: non-monotonic behavior for $z_{tip}$ < 30 nm followed by the decrease at the higher positions; (h) Change of absorption in the apex of the tip and resonant sphere: Absorption in the sphere shows the opposite tendency than absorption in the tip and its effective polarizability.

## IV. Edge resonances: Dipole model and demodulation of absorption in the tip

Approaching the sharp material edge, scattering properties of the s-SNOM tip, and in particular, its effective dipole moment, change. Typically, the edge scan has two features: broad dip (low-contrast line in the image), which always appears outside sample edge of any type of material, and sharp peak (outside bright fringe in the image), which strongly depends on reflective and resonant properties of the edge. The broad decrease of the signal appears because the tip is surrounded by effectively less material at the sharp edge, the effective polarizability decreases, and changes in the scattering are less pronounced. For the ideal sharp edge of the 60-nm thick material, this decrease starts approximately at $x_{tip}$ = 50 nm, and the lowest signal is reached at approximately at $x_{tip}$ = -50... -100 nm.

Based on the outside sharp peak, one can specify two types of edges: Depending on edge shape and material as well as the excitation efficiency of edge resonances, the structure can exhibit either non-resonant or resonant behavior. For non-resonant edges, the signal peak outside the edge always appears at the point where two hot spots are formed: one is below the tip apex and another one is between the tip and vertical side of the material edge (side hot spot). For the case of non-resonant edges, the higher demodulation orders of the s-SNOM signal have less pronounced fringes, which often disappear at the 3$^{rd}$ and 4$^{th}$ demodulation orders. For the side hot spot, the tip-scattered field changes slower than for the hot spot formed at the bottom of the tip on a flat surface. Descending at the sample edge, the tip moves along the vertical side of the edge, and for non-resonant edges, change in the tip position does not bring a large change in the scattered field. As the s-SNOM signal is responsive to the changes in the field, the signal becomes weaker in higher demodulation orders. Therefore, the broad dip in the s-SNOM signal becomes deeper and more pronounced, and the low-contrast line becomes darker for all non-resonant edges. As the outside peak has contributions from two hot spots, it follows the same changes as both of them, and eventually decreases with the increase of demodulation order. Thus, we refer to these edges as non-resonant because the bright fringe is formed always at the position of tip corresponding to two-hot-spots formation and the peak becomes weaker or disappear at higher demodulation orders.

In contrast to the peaks at the non-resonant edge, in the signal of resonant edges, the peaks do not disappear in the higher demodulation orders, and remarkably, the higher demodulation orders have more pronounced peaks. Moving along the vertical side of the sample edge, the tip excites edge resonances, and the efficiency of the excitations changes with the tip position. Thus, the maximum of the signal (fringe) occurs at the tip position with the most efficient excitation of edge resonance. This position may not correspond to the two-hot-spot formation. The edge resonance can have an origin related to either phonon- or plasmon-polariton resonances, Fabry-Perot reflections from sample surfaces (the case of hyperbolic rays in hBN), or any other resonance with the large field increase near the edge.

To analyze near-field properties at the resonant edges of layers, we choose materials with two types of resonances: one is SiC at wavelength $\lambda$ = 11.3 µm having $\varepsilon_m$ = -7 + 0.3i, and another one is hBN with hyperbolic dispersion ($\varepsilon_{in-plane}$ = -20+0.8i, $\varepsilon_{out-of-plane}$ = 2.8 + 8e-4i, at wavelength $\lambda$ = 7.1 µm, data from Ref. [43]). In all cases, the layers are 60-nm thick and on a silicon substrate. Both vertical and horizontal changes of the s-SNOM signal and absorption at the edge of the resonant structures are shown in Fig. 2. The first material supports phonon-polariton resonances, similar to resonances in any complex structures with permittivities close in absolute values but opposite in signs of real parts (e.g. metal and dielectric in visible). The second one is the resonances related to the excitation of phonon-polariton rays, their propagation in the layer at a specific angle defined by the permittivity tensor (see studies in [20,21,44]), and multiple reflections from layer boundaries.

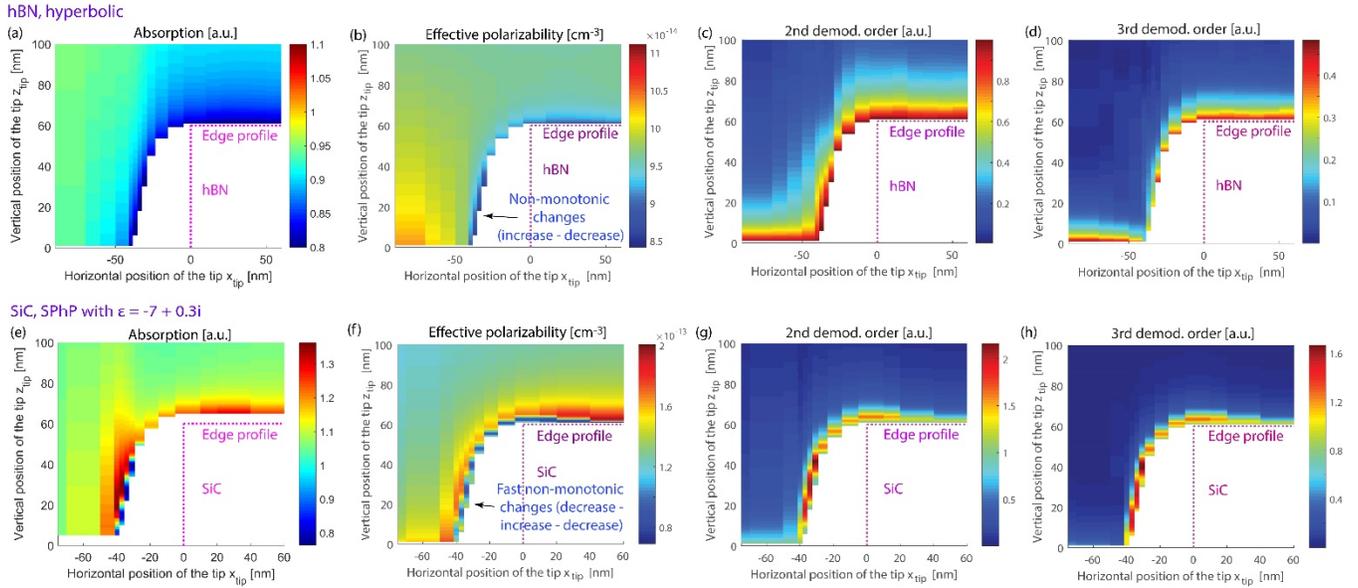

Fig. 2. Vertical and horizontal changes of the s-SNOM signal and absorption at the edge of the resonant structures. (a)-(d) hBN in the hyperbolic regime ($\varepsilon_{\text{in-plane}}$ = -20+0.8i, $\varepsilon_{\text{out-of-plane}}$ = 2.8 + 8e-4i, which is hBN at λ = 7.1 μm); (e)-(h) SiC with $\varepsilon_m$ = -7 + 0.3i. (a),(e) Absorption, (b),(f) Effective polarizability, (c),(g) 2$^{nd}$ demodulation order of s-SNOM signal, (d),(h) 3$^{rd}$ demodulation order of s-SNOM signal. (c)-(d) and (g)-(h) are obtained by demodulating (b) and (f) respectively, and it is a quantity usually observed in s-SNOM measurements in either vertical (approach curve) or horizontal scans. Rapid changes (non-monotonic for SiC) of the signal are seen while the tip moves up/down along the edge. Because of the conical shape of the tip, its trajectory does not coincide with vertical profile of the edge, and for this reason, the white spots in the plots have no signal.

Figure 3a,b represents the case when the tip is next to the surface, and for both materials, we observe bright fringe at the edge. Because of the resonant nature of the fringe (rather than the increased reflection from hot spots), the peaks remain in high demodulation orders and become even more pronounced. Calculations of s-SNOM signal and demodulation harmonics of absorption well agree: positions of the peaks coincide for signal and absorption calculations, and the peaks are always offset from the position of two-hot-spot excitation related to increased reflectance. However, s-SNOM signal calculations through dipole model for SiC give a continuous increase of the peak with an increase of demodulations order, which is most probably an artifact related to dipole approximation. In contrast, demodulation orders of absorption only slightly increase for higher harmonics and thus more appropriate for the case of phonon-polariton edge resonances.

We also compare changes in effective polarizability of the tip with enhancement of the electric field at the point where edge resonance is excited in the most efficient way and outside edge fringes have maxima. Figure 3c shows the vertical scans for hBN at $x_{tip}$ = -31 nm, and Fig. 3d for SiC at $x_{tip}$ = -37 nm. We see that for material with hyperbolic dispersion, the polarizability significantly decreases and the field has a maximum. This behavior is consistent with the analysis of resonance at the flat resonant surface (see Fig. 1). For the SiC edge with resonance, similar to LSPhR in nanosphere, the polarizability and field changes are non-monotonically. Yet, at $z_{tip}$ > 30 nm, all quantities of interest decrease for both materials as the tip is far away from the edge.

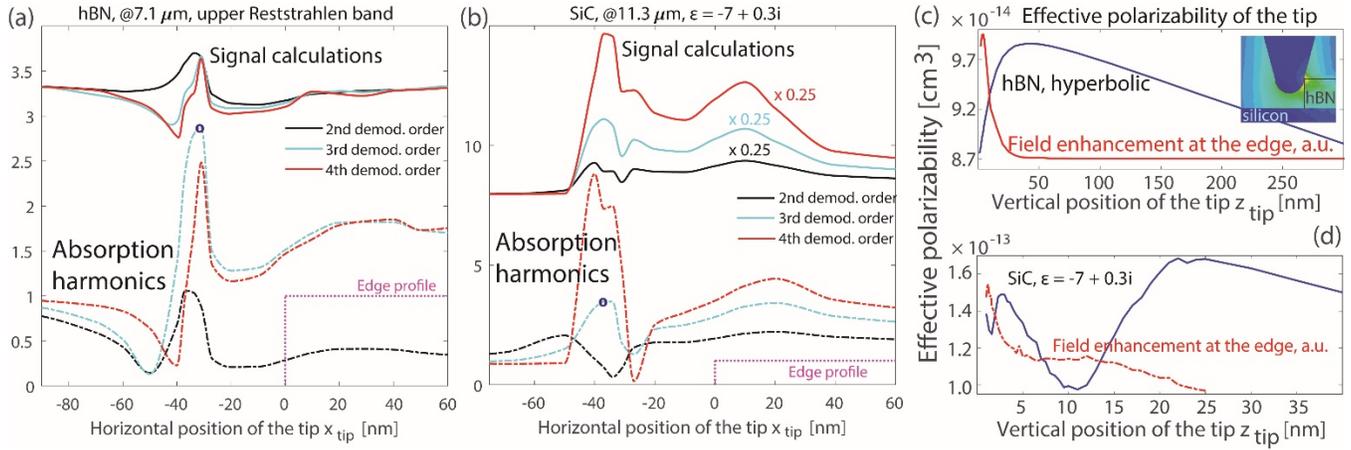

Fig. 3. Signal and absorption harmonics of the s-SNOM scanning the edge of the resonant structures. (a) Harmonics of absorption and s-SNOM signal at the edge of material with hyperbolic dispersion the same as in Fig. 2; (b) Harmonics of absorption and s-SNOM signal at the SiC edge with $\varepsilon_m$ = -7 + 0.3i; In (a) and (b), the signal is normalized so that it is equal to 1 on the substrate at $x_{tip}$ = -2 μm in each demodulation order. In the plots, signal calculations are shifted by 2.5 for hBN and by 7 for SiC for clarity. Edge profile is shown schematically without regard to the ordinate axis, and both hBN and SiC layers are 60-nm thick. The blue circle marks show the coordinate where panels (c) and (d) are calculated ($x_{tip}$ = -31 nm for hBN and $x_{tip}$ = -37 nm for SiC). For both structures, quantitative agreement between signal calculations and harmonics of absorption in the tip means that position of the resonance excitation is adequately captured by the models. (c) Change of effective polarizability of the dipole induced in the tip and the sample at the hBN edge for the tip moving away from the surface. Effective polarizability is decreased at $z_{tip}$ < 20 nm, and the field is significantly enhanced there (the opposite tendency similar to Fig. 1e,h). The plot corresponds to $x_{tip}$ = -31 nm, where the peaks of 3rd and 4th demodulation orders of the hBN edge are. Peak position is offset from the point where two hot spots are formed simultaneously (tip/edge and tip/substrate at $x_{tip}$ = -37 nm). Inset: tip next to the hBN edge at the position of most effective excitation of edge resonance and phonon-polariton rays. (d) The same as (c), but for the edge of SiC. The first 25 nm within the surface is accompanied by the strong increase of the field and non-monotonic changes of the tip effective polarizability (again, the opposite tendency similar to Fig. 1e,h). The plot corresponds to the case $x_{tip}$ = -37 nm, where the peaks of 3rd and 4th demodulation orders of SiC are.

**V. Summary and Conclusion**

Recently emerged layered materials with low-loss phonon-polariton resonances are promising building blocks for subwavelength optical structures, metasurfaces, and functional elements in the infrared range. Scattering-type near-field optical microscopy overcomes the diffraction limit and enables subwavelength characterizations: resolution of s-SNOM is determined by the radius of the tip apex, which is ~30 nm, that is two orders of magnitude less than the mid-infrared wavelength. We have studied properties of the surface and edge resonances in the modeling of s-SNOM measurements. Resonances in the flat surface or nanosphere provide a basic understanding of changes in effective polarizability of the tip, field enhancement at different parts, and energy absorption there. As the first point of the paper, we have shown that excitation of the resonance is accompanied by a decrease of effective polarizability and field in the tip, in contrast to the increased field in the structure.

To study resonances in more complex structures, such as SiC and hBN edges with phonon-polaritons, we have employed two techniques: the first one is calculations of the s-SNOM signal in dipole approximation, and the second one is demodulation of absorption in the tip apex. Calculations of the s-SNOM signal are based on extracting effective polarizability of the tip from reflection coefficient, heavily rely on dipole approximation, and thus more preferable for the structures where the tip has only one hot spot, e.g. flat surface. As the second point of the paper, we have shown that demodulation of absorption should be applied in the case when tip scatters higher multipoles, like at the edge of material that thickness is comparable to tip apex radius. The proposed technique accounts for all hot spots excited around tip apex.

Applicability of both models is studied for edge resonances in a material with phonon-polariton resonances including hBN in mid-infrared with hyperbolic dispersion and SiC. In the material with hyperbolic dispersion, the resonances related to the excitation of phonon-polariton rays and their multiple reflections from layer boundaries. At the tip position that corresponds a maximum of the outside bright fringe and indicates the most efficient excitation of edge resonance, effective polarizability drops within the first several tens of nanometers from the surface, and the field enhancement at the edge strongly increases at the same time. For hBN, both modeling approaches give results that are in good agreement. The SiC edge excites resonances similar to other structures with phonon-polariton or plasmonic resonances. Complex field distribution accompanies these resonances, multiple hot-spots are excited resulting in multipole excitations in the tip, and calculation through absorption demodulation is preferable over the calculation of s-SNOM signal through effective polarizability and dipole approximation.

One should distinguish edge fringes from other fringes that appear at the sample surface because of the propagation of surface and bulk waves in the sample. Often, to specify dispersion of the wave in the sample, the fringe analysis includes spatial Fourier transform [21,34], and in this case, it is important that the outside fringes are excluded from that analysis. Thus, the presence of the outside bright fringes gives information about the material type, and for the case of mixed materials (e.g., different particles on the surface [33]), the presence of bright halo indicates particles with metallic properties.

**Appendix: s-SNOM signal calculations**

We perform full-wave numerical simulations of the structure with realistic shape and permittivity by the frequency-domain solver of CST Microwave Studio. Tip apex is modeled as a hemisphere with radius 30 nm, the whole tip has a conical shape truncated at the apex, the cone base radius is 300 nm, and the cone height is 1 μm. Simulation tests with a smaller height showed that only with the height of 1 μm, the convergence of results can be achieved. The tip is made of silicon and covered with 2 nm platinum layer.

Various structures are placed on the silicon substrate. The effective tip-sample polarizability is $\alpha_{\text{eff}}(h) = i a_1 a_2 (r^{(\text{CST})})^*(h) \exp[-2ik_0 \cos\theta(z_{\max} - z_{\text{pos}})]/(2\pi k_0 \tan\theta)$, where the height-dependent reflectance $r^{(\text{CST})}(h)$ is obtained from numerical modeling, $a_1$ and $a_2$ are the lateral dimensions of the simulation domain, periodic boundary conditions are imposed, $k_0$ is the wave vector of the incident beam, and $\theta$ is the angle of incidence, $z_{\max}$ is the upper boundary of the simulation domain, and $z_{\text{pos}}$ is the coordinate of the substrate or sample surface. The simulations are performed for the whole structure with the tip at different heights starting from 2 nm above the surface and assuming 60-nm oscillation amplitude. The set of points is distributed unevenly to account that most of the changes in effective polarizability occur within the first tens of nanometers from the surface.

The effective polarizability $\alpha_{\text{eff}}(h)$ is used to calculate the far-field scattering given by $E_s(h_0 + \Delta h \sin\omega_T t) = \sum_n s_n e^{in\omega_T t}$, where $s_n$ is the complex s-SNOM signal, $n$ is the demodulation order, $t$ is time, and $\omega_T = f_T / 2\pi$ and $f_T \approx$ 0.3-0.03 MHz is the natural frequency of the probe tip. Further, we sum up far-field radiation $E_S$ with a reference field $E_M$ reflected from an oscillating mirror, to reproduce a pseudoheterodyne interferometer scheme ($f_M \approx$ 0.3 kHz). Finally, we theoretically demodulate the amplitude of the s-SNOM signal and normalize it by dividing the signal at the particular point of the structure to the signal at the substrate far away from the structure: $A_n = |s_{n,\text{sample}}|/|s_{n,\text{substrate}}|$.

**References**

[1] R. Hillenbrand, T. Taubner, F. Keilmann, "Phonon-enhanced light-matter interaction at the nanometre scale," Nature 418, 159 (2002).

[2] S. Dai, Z. Fei, Q. Ma, A.S. Rodin, M. Wagner, A.S. McLeod, M.K. Liu, W. Gannett, W. Regan, K. Watanabe, T. Taniguchi, M. Thiemens, G. Dominguez, A.H. Castro Neto, A. Zettl, F. Keilmann, P. Jarillo-Herrero, M. M. Fogler, D. N. Basov, "Tunable Phonon Polaritons in Atomically Thin van der Waals Crystals of Boron Nitride," Science 343, 1125 (2014).

[3] T. Low, A. Chaves, J.D. Caldwell, A. Kumar, N.X. Fang, P. Avouris, T.F. Heinz, F. Guinea, L. Martin-Moreno, F. Koppens, "Polaritons in layered two-dimensional materials," Nature Materials 16, 182–194 (2017).

[4] J. D. Caldwell, I. Vurgaftman, J. G. Tischler, O. J. Glembocki, J. C. Owrutsky, and T. L. Reinecke, "Atomic-scale photonic hybrids for mid-infrared and terahertz nanophotonics," Nature Nanotechnology 11, 9–15 (2016).

[5] J.D. Caldwell, L. Lindsay, V. Giannini, I. Vurgaftman, T.L. Reinecke, S.A. Maier, O.J. Glembocki, "Low-loss, infrared and terahertz nanophotonics using surface phonon polaritons," Nanophotonics 4, 44 (2015).

[6] T. Taubner, D. Korobkin, Y. Urzhumov, G. Shvets, R. Hillenbrand, "Near-field microscopy through a SiC superlens," Science 313, 1595 (2006).

[7] J.A. Schuller, R. Zia, T. Taubner, M.L. Brongersma, "Dielectric metamaterials based on electric and magnetic resonances of silicon carbide particles," Physical Review Letters 99, 107401 (2007).

[8] J. D. Caldwell, O. J. Glembocki, Y. Francescato, N. Sharac, V. Giannini, F. J. Bezares, J. P. Long, J. C. Owrutsky, I. Vurgaftman, J. G. Tischler, V. D. Wheeler, N. D. Bassim, L.M. Shirey, R. Kasica, and S.A. Maier, "Low-Loss, Extreme Subdiffraction Photon Confinement via Silicon Carbide Localized Surface Phonon Polariton Resonators," Nano Lett. 13, 3690 (2013).

[9] P. Li, X. Yang, T.W.W. Maß, J. Hanss, M. Lewin, A.-K. U. Michel, M. Wuttig, T. Taubner, "Reversible optical switching of highly confined phonon-polaritons with an ultrathin phase-change material," Nature Materials 15, 870 (2016).

[10] A. Kumar, T. Low, K.H. Fung, P. Avouris, and N.X. Fang, "Tunable Light–Matter Interaction and the Role of Hyperbolicity in Graphene–hBN System," Nano Lett. 15(5), 3172–3180 (2015).

[11] P. Li, I. Dolado, F. J. Alfaro-Mozaz, A. Yu. Nikitin, F. Casanova, L. E. Hueso, S. Vélez, and R. Hillenbrand, "Optical Nanoimaging of Hyperbolic Surface Polaritons at the Edges of van der Waals Materials," Nano Lett. 17, 228–235 (2017).

[12] A. Poddubny, I. Iorsh, P. Belov, and Y. Kivshar, "Hyperbolic metamaterials," Nature Photonics 7, 948–957 (2013).

[13] J. Sun, N. M. Litchinitser, and J. Zhou, "Indefinite by Nature: From Ultraviolet to Terahertz," ACS Photonics 1(4), 293–303 (2014).

[14] A.V. Chebykin, V.E. Babicheva, I.V. Iorsh, A.A. Orlov, P.A. Belov, and S.V. Zhukovsky, "Enhancement of the Purcell factor in multiperiodic hyperboliclike metamaterials," Phys. Rev. A 93, 033855 (2016).

[15] S.V. Zhukovsky, A. Orlov, V.E. Babicheva, A.V. Lavrinenko, J. E. Sipe, "Photonic-band-gap engineering for volume plasmon polaritons in multiscale multilayer hyperbolic metamaterials," Phys. Rev. A 90, 013801 (2014).

[16] A.A. Orlov, A.K. Krylova, S.V. Zhukovsky, V.E. Babicheva, P.A. Belov, "Multi-periodicity in plasmonic multilayers: general description and diversity of topologies," Phys. Rev. A 90, 013812 (2014).

[17] A.A. Orlov, E.A. Yankovskaya, S.V. Zhukovsky, V.E. Babicheva, I.V. Iorsh, and P.A. Belov, "Retrieval of Effective Parameters of Subwavelength Periodic Photonic Structures," Crystals 4, 417-426 (2014).

[18] V.Babicheva "Multipole resonances and directional scattering by hyperbolic-media antennas," arXiv:1706.07259, 2017

[19] F.J. Alfaro-Mozaz, P. Alonso-Gonzalez, S. Velez, I. Dolado, M. Autore, S. Mastel, F. Casanova, L.E. Hueso, P. Li, A.Y. Nikitin, R. Hillenbrand, "Nanoimaging of resonating hyperbolic polaritons in linear boron nitride antennas," Nat Commun. 8:15624 (2017).

[20] P. Li, M. Lewin, A. V. Kretinin, J. D. Caldwell, K. S. Novoselov, T. Taniguchi, K. Watanabe, F. Gaussmann, T. Taubner, "Hyperbolic phonon-polaritons in boron nitride for near-field optical imaging and focusing," Nature Commun. 6:7507 (2015).

[21] S. Dai, Q. Ma, T. Andersen, A.S. Mcleod, Z. Fei, M.K. Liu, M. Wagner, K. Watanabe, T. Taniguchi, M. Thiemens, F. Keilmann, P. Jarillo-Herrero, M.M. Fogler, D.N. Basov, "Subdiffractional focusing and guiding of polaritonic rays in a natural hyperbolic material," Nature Commun. 6:6963 (2015).

[22] V. Babicheva, "Hyperbolic-metamaterial waveguides for long-range propagation", arXiv preprint arXiv:1707.07406, 2017

[23] V.E. Babicheva, M.Y. Shalaginov, S. Ishii, A. Boltasseva, and A.V. Kildishev, "Long-range plasmonic waveguides with hyperbolic cladding," Optics Express 23, 31109-31119 (2015).

[24] V.E. Babicheva, M.Y. Shalaginov, S. Ishii, A. Boltasseva, and A.V. Kildishev, "Finite-width plasmonic waveguides with hyperbolic multilayer cladding," Opt. Express 23, 9681-9689 (2015).

[25] S. Ishii, M. Y. Shalaginov, V.E. Babicheva, A. Boltasseva, and A.V. Kildishev, "Plasmonic waveguides cladded by hyperbolic metamaterials," Optics Letters 39, 4663-4666 (2014).

[26] A. Yu. Nikitin, E. Yoxall, M. Schnell, S. Velez, I. Dolado, P. Alonso-Gonzalez, F. Casanova, L. E. Hueso, R. Hillenbrand, "Nanofocusing of Hyperbolic Phonon Polaritons in a Tapered Boron Nitride Slab," ACS Photonics 3, 924 (2016).

[27] J. M. Stiegler, Y. Abate, A. Cvitkovic, Y.E. Romanyuk, A.J. Huber, S.R. Leone, R. Hillenbrand, "Nanoscale Infrared Absorption Spectroscopy of Individual Nanoparticles Enabled by Scattering-Type Near-Field Microscopy," ACS Nano 5, 6494–6499 (2011).

[28] A.A. Govyadinov, S. Mastel, F. Golmar, A. Chuvilin, P.S. Carney, and R. Hillenbrand, "Recovery of Permittivity and Depth from Near-Field Data as a Step toward Infrared Nanotomography," ACS Nano 8, 6911 (2014).

[29] Y. Abate, S. Gamage, L. Zhen, S. B. Cronin, H. Wang, V. Babicheva, M. H. Javani, and M. I. Stockman, "Nanoscopy Reveals Metallic Black Phosphorus," Light Science Applications, 5, e16162 (2016).

[30] V.E. Babicheva, S. Gamage, M.I. Stockman, Y. Abate, "Near-field edge fringes at sharp material boundaries," Optics Express 2017, preprint arXiv:1707.08686.

[31] Y. Abate, D. Seidlitz, A. Fali, S. Gamage, V. E. Babicheva, V. S. Yakovlev, M. I. Stockman, R. Collazo, D. E. Alden, and N. Dietz, "Nanoscopy of Phase Separation in $In_{1-x}Ga_xN$ Alloys," ACS Appl. Mater. Interfaces, 8(35), 23160 (2016).

[32] Y. Abate, V.E. Babicheva, V.S. Yakovlev, and N. Dietz, "Towards Understanding and Control of Nanoscale Phase Segregation in Indium-Gallium-Nitride Alloys," pp. 183-207, Chapter 6 in "III-Nitride Materials, Devices, and Nano-Structures," 424 pages, Ed: Zhe Chuan Feng, World Scientific Publishing, 2017.

[33] A. Cvitkovic, N. Ocelic, and R. Hillenbrand, "Material-Specific Infrared Recognition of Single Sub-10 nm Particles by Substrate-Enhanced Scattering-Type Near-Field Microscopy," Nano Lett. 7, 3177-3181 (2007).

[34] V. E. Babicheva, S. Gamage, L. Zhen, S. B. Cronin, V. S. Yakovlev, Y. Abate, "Near-field Surface Waves in Few-Layer $MoS_2$," preprint arXiv:1707.07743, 2017

[35] A. Woessner, M.B. Lundeberg, Y. Gao, A. Principi, P. Alonso-González, M. Carrega, K. Watanabe, T. Taniguchi, G. Vignale, M. Polini, J. Hone, R. Hillenbrand, F.H.L. Koppens, "Highly confined low-loss plasmons in graphene–boron nitride heterostructures," Nature Materials 14, 421–425 (2015).

[36] S. Dai, Q. Ma, M. K. Liu, T. Andersen, Z. Fei, M. D. Goldflam, M. Wagner, K. Watanabe, T. Taniguchi, M. Thiemens, F. Keilmann, G. C. A. M. Janssen, S-E. Zhu, P. Jarillo-Herrero, M. M. Fogler, D. N. Basov, "Graphene on hexagonal boron nitride as a tunable hyperbolic metamaterial," Nature Nanotechnology 10, 682–686 (2015).

[37] A. S. McLeod, P. Kelly, M. D. Goldflam, Z. Gainsforth, A. J. Westphal, G. Dominguez, M. H. Thiemens, M. M. Fogler, and D. N. Basov, "Model for quantitative tip-enhanced spectroscopy and the extraction of nanoscale-resolved optical constants," Phys. Rev. B 90, 085136 (2014).

[38] L. M. Zhang, G. O. Andreev, Z. Fei, A. S. McLeod, G. Dominguez, M. Thiemens, A. H. Castro-Neto, D. N. Basov, and M. M. Fogler, "Near-field spectroscopy of silicon dioxide thin films," Phys. Rev. B 85, 075419 (2012).

[39] T. Neuman, P. Alonso-Gonzalez, A. Garcia-Etxarri, M. Schnell, R. Hillenbrand, and J. Aizpurua, "Mapping the near fields of plasmonic nanoantennas by scattering-type scanning near-field optical microscopy," Laser Photonics Rev. 9, 637–649 (2015).

[40] B.-Y. Jiang, L. M. Zhang, A. H. Castro Neto, D. N. Basov, and M. M. Fogler, "Generalized spectral method for near-field optical microscopy," Journal of Applied Physics 119, 054305 (2016).

[41] Le Wang and Xiaoji G. Xu, "Scattering-type scanning near-field optical microscopy with reconstruction of vertical interaction," Nature Commun. 6:8973 (2015).

[42] S. Babar, J. H. Weaver, "Optical constants of Cu, Ag, and Au revisited," Appl. Opt. 54, 477-481 (2015).

[43] Y. Cai, L. Zhang, Q. Zeng, L. Cheng, Y. Xu, "Infrared reflectance spectrum of BN calculated from first principles," Solid State Communications 141, 262–266 (2007).

[44] S. Ishii, A. V. Kildishev, E. Narimanov, V. M. Shalaev, V. P. Drachev, "Sub-wavelength interference pattern from volume plasmon polaritons in a hyperbolic medium," Laser Photonics Rev. 7, 265–271 (2013).